\documentclass[conference]{IEEEtran}
\newtheorem{definition}{Definition}
\IEEEoverridecommandlockouts
\usepackage{cite}
\usepackage{subfigure}
\usepackage{amsmath,amssymb,amsfonts}
\usepackage{algorithmic}
\usepackage{graphicx}
\usepackage{textcomp}
\usepackage{xcolor}
\usepackage{xspace}
\usepackage{comment}
\def\BibTeX{{\rm B\kern-.05em{\sc i\kern-.025em b}\kern-.08em
    T\kern-.1667em\lower.7ex\hbox{E}\kern-.125emX}}
\begin{document}

\title{ IoTC$^2$\xspace : A Formal Method Approach for Detecting Conflicts in Large Scale IoT Systems}

\newcommand{\system}{IoTC$^2$\xspace}

\author{
    \IEEEauthorblockN{Abdullah Al Farooq\IEEEauthorrefmark{1}, Ehab Al-Shaer\IEEEauthorrefmark{1}, Thomas Moyer\IEEEauthorrefmark{1}, Krishna Kant\IEEEauthorrefmark{2}}
    \IEEEauthorblockA{\IEEEauthorrefmark{1}The University of North Carolina at Charlotte
    \\\{afarooq,ealshaer,tom.moyer\}@uncc.edu}
    \IEEEauthorblockA{\IEEEauthorrefmark{2}Temple University
    \\\{kkant\}@temple.edu}
}

\maketitle

\begin{abstract}
Internet of Things (IoT) has become a common paradigm for different domains such as health care, transportation infrastructure, smart home, smart shopping, and e-commerce. With its interoperable functionality, it is now possible to connect all domains of IoT together for providing competent services to the users. Because numerous IoT devices can connect and communicate at the same time, there can be events that trigger conflicting actions to an actuator or an environmental feature. However, there have been very few research efforts made to detect conflicting situation in IoT system using formal method. This paper provides a formal method approach, IoT Confict Checker (\system), to ensure safety of controller and actuators' behavior with respect to conflicts. Any policy violation results in detection of the conflicts. We defined the safety policies for controller, actions, and triggering events and implemented the those with Prolog to prove the logical completeness and soundness. In addition to that, we have implemented the detection policies in Matlab Simulink Environment with its built-in Model Verification blocks. We created smart home environment in Simulink and showed how the conflicts affect actions and corresponding features. We have also experimented the scalability, efficiency, and accuracy of our method in the simulated environment.
\end{abstract}


\section{Introduction}
Recently, the Internet of Things (IoT) has become a hot topic in the technology community. The deployment and application of IoT devices covers a variety of fields, ranging from the smart homes to transportation management system. It is projected that there will be as many as 50 billion connected devices by 2020 ~\cite{dave2011next}. IoT presents a unique advantage in that is connects devices within and across domains, e.g. smart homes, traffic route guidance systems, shopping systems, ride sharing, and parking management systems. IoT applications and devices share required data and provide unified services to the users.

However, the distributed nature of IoT leaves devices and communication channels exposed to attackers and many of these devices and protocols are resource-constrained. This, combined with the fact that many of these devices receive infrequent updates leaves them highly susceptible to attack. An attacker can trigger an event that leads to conflicting actions for the same object or feature of the environment. As for example, an attacker can create multiple events that trigger a thermostat to increase and decrease temperature of a room at the same time. Sending two different commands in the thermostat at the same time continuously can damage it, by artificially shortening the devices lifespan. In this way, the attacker not only damages an asset, but also may drive the occupants of the room to leave due to fluctuations in the comfort level of the room. Moreover, misconfiguration is possible as there are numerous rules or policies for taking actions by the controllers after events have occurred.

Attackers can leverage these conflicts and vulnerabilities to gain physical access to a smart building. For example, an attacker may compromise a carbon monoxide sensor and falsely trigger the alarm indicating the presence of carbon monoxide, which in turn sends a command to the smart windows to open allowing fresh air into the building. Occupants may confuse the CO alarm with a fire alarm and leave the building. A thief can target the opened windows  to enter the building. In addition to the attacks described above, an attacker can create series of attacks (cascading attack)~\cite{IoTDDOS}. Even with IoT technology in the early stages of development and deployment, the guarantees of maintaining safe and secure operation of an environment through IoT devices can attract more users. Even legacy, or dumb devices can be attached to the system and be operated through a controller.

Due to the limited computational and memory capacities of IoT devices, it is not always possible to secure and monitor each and every device and communication channel. A controller or a group of controllers provide the computational and storage capacity to make decisions based on events coming from the edge devices (i.e.  sensors) and issue commands to the appropriate actuators. The automated decisions made by a controller may try to command a device which is already performing a different action. Moreover, devices will be connecting to the IoT controller intermittently. Events can occur any time or an attacker can force specific events to happen in order to take advantage of misconfigurations (i.e. rule conflicts) in the IoT system. Furthermore, the operational policies for an IoT system can change over time based on the requirements of the building and its occupants. Hence, it is important to check whether a recently triggered event causes a conflict based on the ruleset that is stored in the controllers.

In this research, we propose \system, a formal methods approach to ensure safety properties for the controllers and actuators in an IoT system. 
The main contributions of the paper are:
\begin{itemize}
\item a formal approach to defining the safety properties of an IoT system
\item a technique for detecting conflicts within the rulesets of the IoT system that violate the safety properties of the system
\item an implementation of \system that can be used in real-time to ensure the safety of the IoT system
\end{itemize}
It should be noted that the actuation commands are issued from the controllers. Hence, multiple controllers that try to command the same actuator are in scope for this work. \system has all the rules/logic for the IoT system so that when events trigger a rule or a set of rules, \system makes sure the safety properties are maintained. When a violation occurs of the safety properties occurs it is due to conflicts in the rules defined for the system. It is also worth noting that the developed framework is capable of detecting conflicts that are caused by misconfiguration of operational rules. The conflicts are detected even before the conflicting actions take place. \system is capable identifying the specific type of violation that has occurred. We have also implemented an IoT environment simulation in Matlab's Simulink environment to understand the impact of conflicts in an IoT system.

\section{\system Framework}
A basic IoT system is comprised of a number of sensors, actuators, and controllers. The sensors collect measurements of the current state of the environment or the system. There are operational rules stored in the controller. These rules are triggered based on events collected by the sensors. The rules define what actions should be taken by the various actuators connected to the controller. Simply put, an event triggers a rule and the rule triggers an action in an IoT system. The controller decides what action or set of actions to take, as defined by the rules installed in the controller. The controller issues commands to the actuator for performing the most appropriate action. The devices are connected through wired or wireless networks.

\begin{figure}[htbp]
\centering
\hspace*{-.5cm}     
\includegraphics[width=0.95\linewidth, keepaspectratio=true]{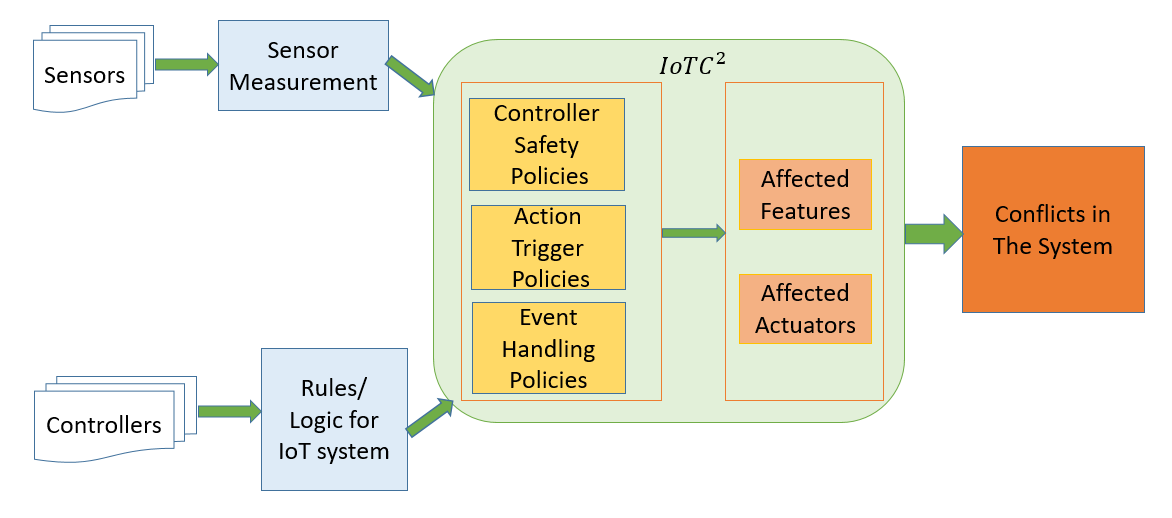}
\vspace{-8pt}
\caption{\system Framework for Conflict Detection}
\label{Fig_Architecture}
\vspace{-8pt}
\end{figure}

The architecture of the \system is given in figure \ref{Fig_Architecture}. All of the rules/logic used for operation in each controller of the IoT system are the input for \system. Whenever there is a change in any rule or an additional rule is added to a controller, its inputted to \system. The second type of inputs for our framework is the sensor measurement with timestamps. To start, our framework receives copies of the traffic from the sensors to the controllers. As soon as \system receives the sensor measurement, by using the rules/logic from the controller, \system determines what actions the controller will emit for the actuators. Next, \system determines the list of  actuators, affected features, and issuing controllers. Using these lists, \system determines whether or not these activities violate the safety properties (or create conflicts) within the IoT system. \system has the capability to output the number of conflicts and their type in the IoT system. In addition to conflict creating events, misconfiguration of rules within an IoT system can lead to a set of commands that can violate the safety properties of the system which \system can also detect.

\section{Formal Method for Detecting Conflicts}
In this section, we present our formal method for detecting safety property violations in IoT systems. In the following description, we rely on the notation defined in Table~\ref{tab:notation}. We start by defining some general properties of the IoT system model, events, triggers, and actions.

\bgroup
\def\arraystretch{1.5}
\begin{table}[ht]
\centering
\begin{tabular}{|l|l|}
\hline
\textbf{Notation}   & \textbf{Explanation}                                                                                     \\ \hline
$e^t_i$             & Event $i$ generated at time $t$                               \\ \hline
$a^{l,f}_{m,n} (t)$ & Actuator $m$ taking action $n$ on feature $f$ at location $l$ \\ \hline
$c^{p}$             & Controller $p$                                                \\ \hline
$x'$                & Object $x'$ can be the same or different from object $x$      \\ \hline
$\bar{x}$           & Object $\bar{x}$ must be different from $x$                   \\ \hline
\end{tabular}
\vspace{1em}
\caption{Notation used\label{tab:notation}}
\end{table}
\egroup

\begin{definition}
If there are features $f_x$ and $f_y$ such that changes in $f_x$ affect $f_y$, then these features are dependent. It can be the case that feature $f_x$ and feature $f_y$ are not directly dependent, however feature $f_x$ affects $f_z$ and $f_z$ affects $f_y$. In this case, $f_y$ is indirectly dependent on $f_x$. Regardless, they are noted as:

\begin{equation}
f_x \overset{d}{=} f_y
\end{equation}
\end{definition}

\begin{definition}
If there are two events that are the same or similar by their characteristics and functionality and they occur within a bounded time frame, these events are called overlapping events. They are noted as:
\begin{equation}
e_1 \overset{o}{=} e_2
\end{equation}
Whenever two events are not overlapping, they are considered disjoint events.
\end{definition}

\subsection{Controller Safety Policies}
The controller is a crucial component of an IoT system that receives measurements from sensors and based on those measurements, it generates actuation commands for the appropriate actuators. We define the controller safety policies as follows:

\begin{itemize}
\item There are no two rules where two or more controllers can trigger the same actuator at the same time.
\begin{equation}
C_1 \colon
\neg((e^t_i \Rightarrow a^{l,f}_{m \in c_p,n}) \wedge (e^t_{i'} \Rightarrow a^{l,f'}_{m \in c_{\bar{p}},n'}))
\end{equation}
An IoT system has number of rules for operating where the same actuator $m$ is controlled by more than one controller $c_p$ and $c_{\bar{p}}$. If the same actuator is accessed at the same time $t$, a conflict occurs. The actions (denoted by subscript $n$ and $n'$) on the actuators can be same or different, which does not change this policy. The affected features in this case are made different (superscript $f$ and $f'$) because the difference does not impact the safety policy. The impact of the potentially different features in creating conflicts is discussed in Section~\ref{sec:eval}. The following are examples of conflict scenarios that can be captured by this rule.
  \begin{itemize}
  \item A smoke detector and a water-leak detector can each trigger an alarm. We assume that the smoke detector and water-leak detectors are controlled by different controllers. But if the same alarm sounds at the same time, it will be hard to distinguish which event triggered the alarm. The policy formalized here restricts multiple controllers from triggering the same action at the same time. 
  \item Motion detected inside an elevator triggers the controller to prevent the elevator door from locking. On the other hand, an alarm in the building will lock the door of the elevator so that no one can use it during the alarm. Here, the door of the elevator is the actuator while the action on it is operated by two controllers.
  \end{itemize}
  
\item There are no two rules where two controllers can trigger actions that affect the same, or dependent, features at the same time.

\begin{equation}
\begin{aligned}
& C_2 \colon \neg((e^t_i \Rightarrow a^{l,f}_{m \in c_p,n}) \wedge (e^t_{i'} \Rightarrow a^{l,f'}_{m' \in c_{\bar{p}},n'})\\
&\wedge (f = f' \vee f \overset{d}{=}f')) 
\end{aligned}
\end{equation}

There can be more than one rule that can trigger different actions (the subscript $m$ and $m'$ denote different actuators). However, these different actuators can impact the same, or dependent features. The same features are identified with '$=$' . On the other hand, the notion $ \overset{d}{=}$ denotes the dependency among two features $f$ and $f'$. An example of such violation a is:

  \begin{itemize}
  \item A window opener and a thermostat can be two different actuators controlled by different controllers, but can affect the same feature (temperature) of the room.
  \end{itemize}

\end{itemize}

\subsection{Multiple Action Trigger Policies}
When an actuator is issued commands to perform multiple actions at the same time, conflicts can occur. In order to prevent conflicts, we have the following safety property:

\begin{itemize}
\item There are no two rules where two or more overlapping events (from any sensor) can trigger multiple actions on the same actuator.
\begin{itemize}
\item Different action $n'$ 
\item Opposite action $\bar{n}$
\item Dependent action $\hat{n}$
\end{itemize}

\begin{equation}
\begin{aligned}
C_3 \colon \neg((e^t_i \Rightarrow a^{l,f}_{m,n}) \wedge (e^t_{i'} \Rightarrow (a^{l,f'}_{m ,\hat{n}} \wedge a^{l,f'}_{m ,n^*} \wedge a^{l,f'}_{m ,\bar{\bar n}}))\\
\wedge (i \overset{o}{=} i'))
\end{aligned}
\label{multAction}
\end{equation}

In this formula, two events are differentiated by the subscript $i$ and $i'$. The measurement of a sensor can trigger an event depending on whether that measurement is 'more than' or 'less than' or 'equal to' a pre-specified value. We define overlapping events as two or more events that occur within a specific time frame, or are related by their signature. The overlapping relation between two events $i$ and $i'$ is denoted by $i \overset{o}{=} i'$. These overlapping events can trigger different actions (e.g. increase temperature or decrease temperature), opposite action (e.g. open the door and close the door), dependent action (beeping and flashing light on an alarm) and the same but overlapping action (increase temperature on a thermostat twice within 5 second). An action $n$ on actuator $m$ is the reference action and any action ($n'$, $\bar{n}$, or $\hat{n}$) other than $n$ is considered as the conflicting action on the same actuator $m$. An example of this safety policy violation is given below:

\begin{itemize}

\item Both room one and room two have temperature sensors but no thermostat. The corridor that joins both room has a thermostat, but no sensor. Temperature decreases in room one and temperature increases in room two can trigger the same thermostat. Hence, the thermostat can be triggered to increase the temperature and decrease the temperature at the same time.
\end{itemize}

\item There are no two rules where overlapping events can trigger actions that affect the same or dependent features.
\begin{equation}
\begin{aligned}
C_4 \colon \neg((e^t_i \Rightarrow a^{l,f}_{m,n}) \wedge (e^t_{i'} \Rightarrow a^{l,f'}_{m ,\bar{n}}) \\
\wedge (i \overset{o}{=} i') \wedge ((f= f') \vee (f \overset{d}{=} f')))
\end{aligned}
\end{equation}

When an action is performed by an IoT device, it may affect one or more features (e.g. temperature, humidity, or luminance). There are some features the actuators affect directly while some features are impacted indirectly. As an example, humidity and temperature can be considered as dependent features~\cite{de2002thermal}. When the temperature goes up, it affects the humidity of a room if no moisture is added. This is because warm air can hold more water vapor than cool air. We differentiate two features by $f$ and $f'$. There can be two rules that get activated at the same time by two overlapping events, resulting in two different actions, $n$ and $\bar{n}$. These two actions affect features $f$ and $f'$ which are either the same or dependent. The following are examples of conflicts when this safety policy is violated:
\begin{itemize}
\item Room one and room two have a shared thermostat in room one. Hence, room one gets hotter than room two when the thermostat is turned on. Based on the temperature reading from room one, the thermostat is asked to turn off. However, the temperature measurement from room two will ask the controller to turn on the thermostat again. As mentioned earlier, temperature and humidity are dependent and therefore it is possible that humidity in two room will vary. If both rooms share a humidifier that is placed in room 2, the overlapping events can turn the humidifier on or off simultaneously. Apart from depending environment features being affected, more energy is needed for additional actuations.

\item Luminance level can be impacted by overlapping events. Window blind and room lights are two different actuators that impact luminance.

\end{itemize}

\item No two or more completely disjoint events can trigger multiple action on the same actuators
\begin{equation}
\begin{aligned}
C_5 \colon \neg((e^t_i \Rightarrow a^{l,f}_{m,n}) \wedge (e^t_{i'} \Rightarrow (a^{l,f'}_{m ,\hat{n}} \vee a^{l,f'}_{m ,n^*} \vee a^{l,f'}_{m ,\bar{\bar n}}))\\
\wedge \neg(i \overset{o}{=} i')))
\end{aligned}
\end{equation}

In a large IoT system, it is not easy to distinguish overlapping events. Hence, we turn our attention to modeling the safety properties that are based on disjoint events. It is possible that these disjoint events are overlooked when devising the IoT operational rules, yet these rules can create conflicts within a single actuator. Examples for such conflicts are:

\begin{itemize}
\item Management can impose a rule stating that when it is after 6 pm, the temperature need not be controlled. This means that the temperature of a smart building will follow the basic thermal model of the building. On the other hand, movement in a room will trigger the thermostat to increase/decrease the temperature for better occupant comfort.
\item Smoke detection and carbon monoxide detection can be two completely disjoint event that trigger multiple alarms to sound at the same time.

\end{itemize}

\item No two completely disjoint events can trigger multiple actions that affect the same, or dependent features.
\begin{equation}
\begin{aligned}
C_6 \colon \neg((e^t_i \Rightarrow a^{l,f}_{m,n}) \wedge (e^t_{i'} \Rightarrow a^{l,f'}_{m ,\bar{n}}) \wedge \neg(i \overset{o}{=} i')\\
\wedge ((f= f') \vee (f \overset{d}{=} f')))
\end{aligned}
\end{equation}

Two rules can be triggered by completely disjoint events at the same time. The actuator and its location are kept unchanged by notation $m$ and $l$, respectively. The actions are differentiated by $n$ and $n'$. At the same location and with the same actuator, two features $f$ and $f'$ got affected. When these two features are dependent, $f \overset{d}{=} f'$ will return true.

\begin{itemize}
\item A window opening or closing and a thermostat turning on or off are two completely disjoint events that impact the temperature of the room
\item Management can impose a rule that when it is after 6 pm, the thermostat and the humidifier should not be adjusted. On the other hand, movement in a room will trigger the thermostat to increase/decrease the temperature for better comfort. This has affect on humidity as these two features are dependent.

\end{itemize}

\end{itemize}

\subsection{Multiple Event Handling Policies}
A controller actuates an IoT device when an event occurs. There is relatively little control over how and when an event is generated. However, the way multiple events are handled, can be controlled. Therefore, we focus on formalizing event handling policies. The formalization is as follows:

\begin{itemize}
\item No single sensor with single objective can create more than one event within a specific time limit.

\begin{equation}
\label{eventHandler1}
\begin{aligned}
C_7 \colon \neg((e^t_i \Rightarrow a^{l,f}_{m,n}) \wedge (e^{\bar{t}}_{j} \Rightarrow a^{l,f}_{m ,n}))
\end{aligned}
\end{equation}

Here, two events $i$ and $i'$ are prohibited from same sensor $j$ within a time limit $t'$. A sensor might send the same measurement to the controller more than once due to any physical or communication issue. This should be handled in a proper way so that same actions are not taken by the same actuator. 

A sensor can send same temperature measurement (e.g. 60F) twice to the controller within a 30 second interval. The controller would instruct the thermostat to increase the temperature by 10F each time it receives the input from the sensor. Therefore, the temperature of the room is increased to 80F. 
\end{itemize}

\subsection{Completeness of IoT Safety Properties}
\begin{definition}
If an IoT system, comprised of sensors  $s_{1...m}$, controllers $cntrl_{1...n}$, and actuators $a_{1...o}$, violates any safety properties $c_{1...p}$, a conflict $c^* \in Conflict$ has occurred.
\end{definition}

Completeness means that you can prove anything that's true. In order to analyze the safety policies formalized above in terms of controllers, triggered actions, and event handling, the policies were implemented using Prolog. If there exists a conflict $c^* \in Conflict$ in the IoT system operations, \system finds it using the backward chaining. Prolog querey evaluation employs \emph{Selective Linear Definite-clause with Negation as Failure} SLDNF~\cite{TriskaProlog}. However, the Dept First Search (DFS) strategy of Prolog makes it logically incomplete. With this strategy, the search begins from one node and traverses a single path to find the query answer, i.e. looking for conflicts. Whenever a conflict is found in the search space Prolog does not traverse that branch to find another conflict even if another one exists.

We followed the way proposed by~\cite{wielemaker2012swi} where the built-in dept-first search of Prolog was overruled. Rather, the implementation was based on iterative deepening of the query where each recursive call takes place at the top level of a conditional. In doing so, we have used tail-recursion. In each recursive call, the number of actions triggered, or the controllers associated with it, or the events that triggered the actions are stored in lists $list_A$, $list_C$, and $list_E$, respectively. In this way, the search space is being completed in lists. Then this list is traversed to find the conflicts in the system. Whenever, a resolution refutation is found, our Prolog implementation finds it and adds 0 (zero) to the accumulator rather than stopping the search process on that node.

The iterative deepening strategy is potentially inefficient. One could argue that the size of lists will grow unbounded. However, we point out that the formulation for conflicts is dependent on the specific time the event has occurred. Say two actions $a_1$ and $a_2$ are triggered at the same actuator $x$ at time $t_1$ and $t_2$. A conflict $C$ occurs if and only if $t_1 = t_2$. These two actions cannot create conflicts on the same actuator if they are triggered at different times. If there are total of $m$ automation/operational rules in all the controllers of an IoT system, there can be at most $n$ rules that are triggered at time $t$ and it is obvious that $n<<m$.

\subsection{Soundness of IoT Safety Properties}

\begin{definition}
\label{soundnessDefinition}
The safety properties of \system is sound, if for all sensors $s_{1...m}$, controllers $cntrl_{1...n}$, and actuators $a_{1...o}$, all possible operations in the system are subset of the authorized operations allowed by \system.
\end{definition}

With the definition of soundness from~\ref{soundnessDefinition}, we can conclude that all safety properties, expressed in conjunctive normal form (CNF) make it logically sound as given in~\ref{SoundnessEq}.

\begin{equation}
\label{SoundnessEq}
\begin{aligned}
C_f \colon C_1 \wedge C_2 \wedge C_3 \wedge C_4 \wedge C_5 \wedge C_6 \wedge C_7
\end{aligned}
\end{equation}

If there exists a conflict in the IoT system, yet \system cannot detect it, we call it unsound. As mentioned earlier, \system is implemented in Prolog where it backtracks till it finds the ground truth. The only way \system can fail, is that the ground truth is corrupted or altered which is left out of scope of this paper. If there exists any resolution refutation, our implementation must find it because of the lists $list_A$, $list_C$, and $list_E$ used to avoid built-in DFS of Prolog programs.

If the soundness and completeness conditions fail for \system, the negation of $C_f$ in~\ref{SoundnessEq} will provide us an example of unsafe situation of the IoT system. More simply a conflict has occurred.

\section{Evaluation\label{sec:eval}}

For out evaluation, we created an IoT environment using Matlab's Simulink. We designed a house with three rooms with corridors attaching each of the rooms. The thermal model of the house was adapted from~\cite{ThermalSimulink}. The rooms have facilities for smoke detection, carbon monoxide detection, smart lights, smart window shutters and blinds, smart doors, etc. Operational rules for automating these Iot devices are implemented using appropriate Simulink blocks. We run each simulation several times to observe whether \system can detect conflicts in the smart house successfully. We have used the built-in model verification blocks (e.g., Assertion, Check Dynamic Gap, Check Dynamic Range, etc) of Simulink to help our detection. \system outputs the total count of each different type of detected conflict in a given simulation step. Not only are the detected conflicts shown, but also their effects on features or actuators.

\begin{figure}[htbp]
\centering
\includegraphics[scale=0.34, keepaspectratio=true]{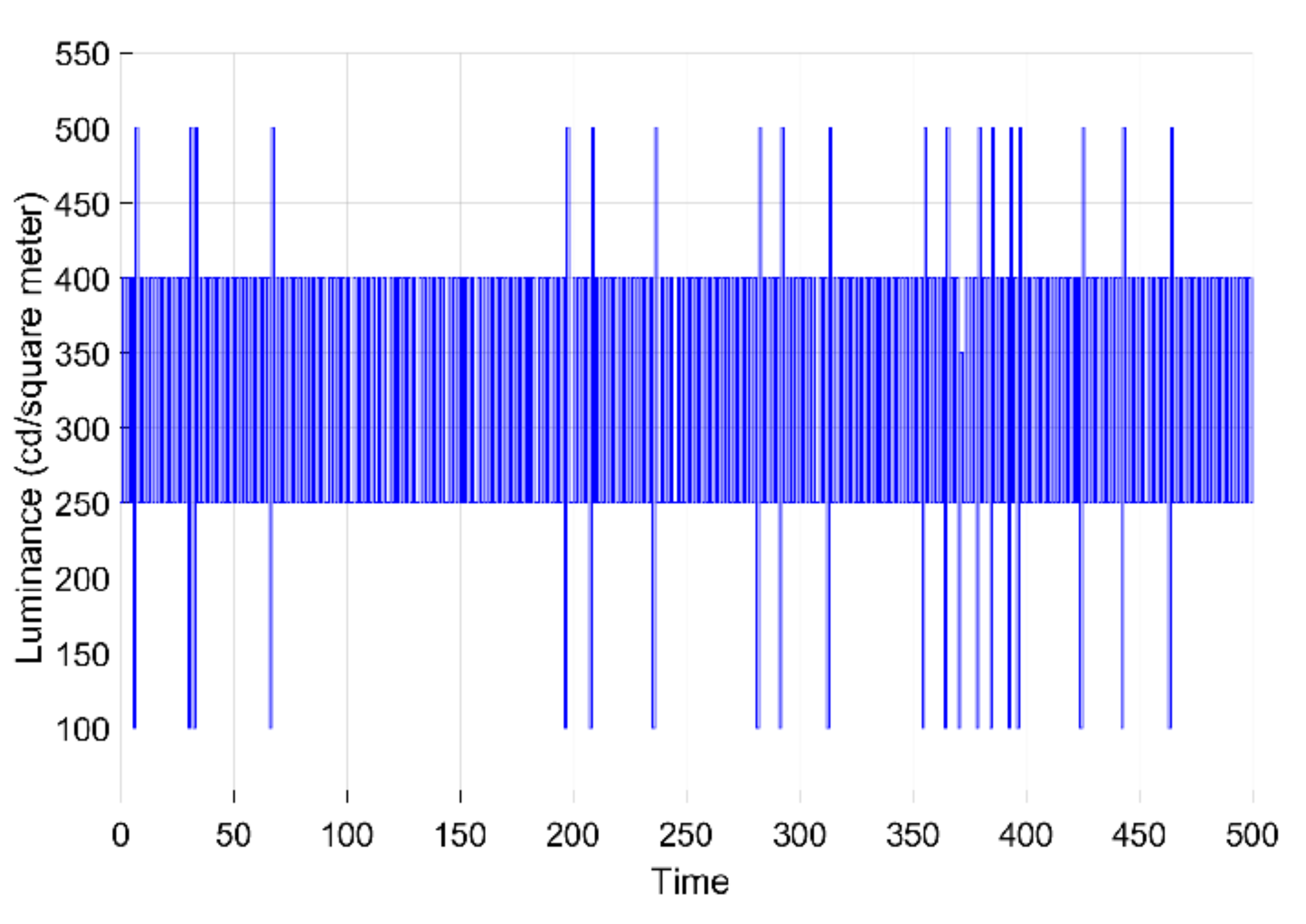}
\vspace{-9pt}
\caption{Luminance range of a room when smart window blinder and smart light are considered}
\label{A1_Difference_Luminance}
\vspace{-9pt}
\end{figure}

First, we evaluate the effects of conflicts on actuators or the environment feature. Later, we discuss how well \system can detect the conflict in a simulated environment. Our first experiment was conducted to observe whether two different actions on different actuators affect the same environmental feature. A smart window shutter is sometimes operated by the occupant with the help of an app from tablet or smart phone. Similarly, the shutter is operated to keep open at different times of the day. On the other hand, the smart home is automated to turn the smart lights on whenever it finds movement in a room. In our setup, the probability of both the opening up the window shutter and turning on smart light was set as 0.10. The conflict occurs during a split second of time when both window shutter and smart light turns on or off together given that someone is in the room. The luminance of the room then gets out of the range ($>450$ or $<200$) compared to a comfortable luminance range. The simulation was run for 500 units of time. We see the the experimental observations in Figure~\ref{A1_Difference_Luminance}. As can be seen from the figure, the luminance of the room exceeds the set bounds for a comfortable luminance level.

\begin{figure}[htbp]
\centering
\includegraphics[scale=0.34, keepaspectratio=true]{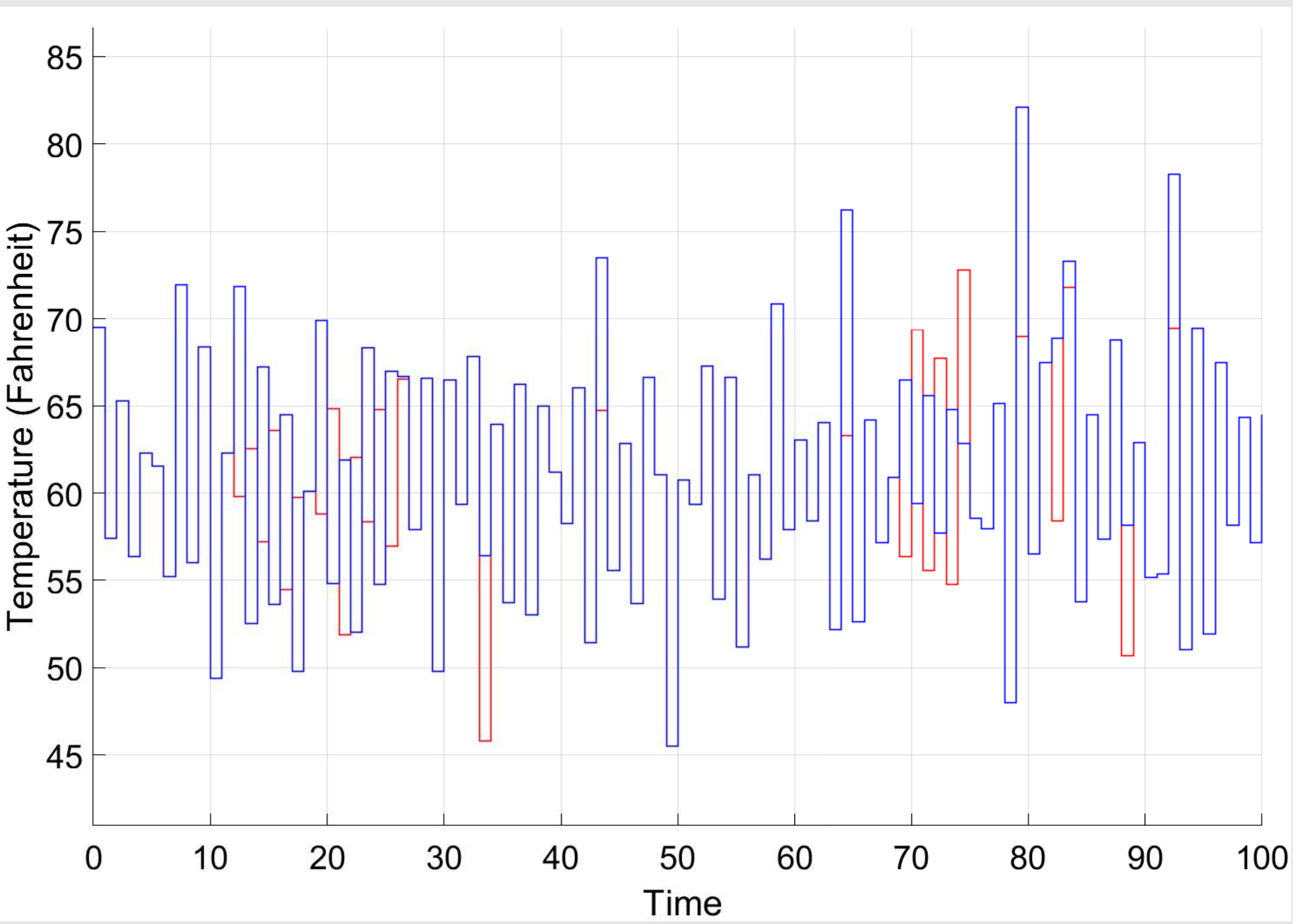}
\vspace{-9pt}
\caption{Effect on Temperature when thermostat and window shutter works at the same time}
\label{C1_Difference}
\vspace{-9pt}
\end{figure}

\begin{figure*}[htbp] 
\vspace{-9pt}   
    \begin{center}                               
    	\subfigure[]{
           \label{C1Count}
           \includegraphics[scale=0.3,keepaspectratio=true]{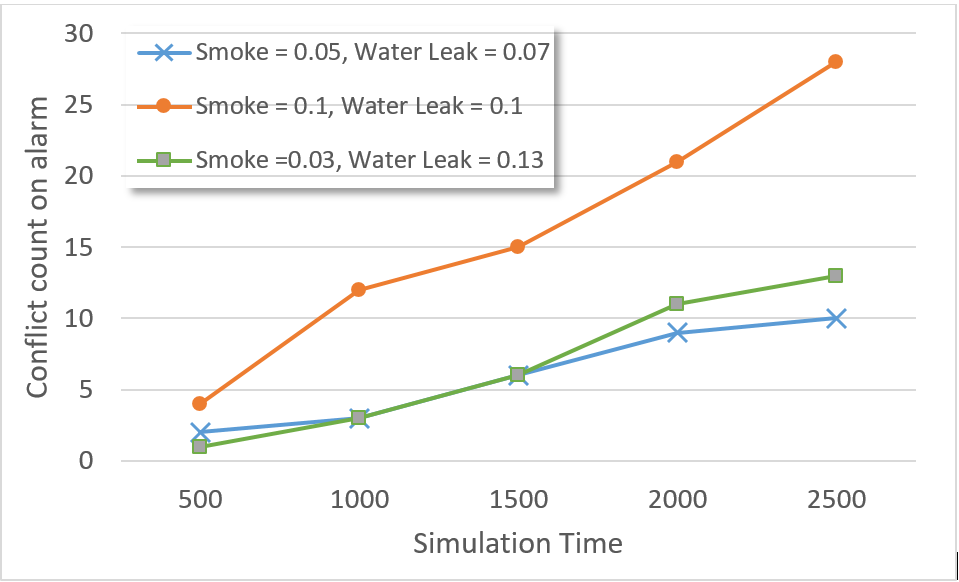} 
        }
    	\subfigure[]{
           \label{C2Count}
           \includegraphics[scale=0.3,keepaspectratio=true]{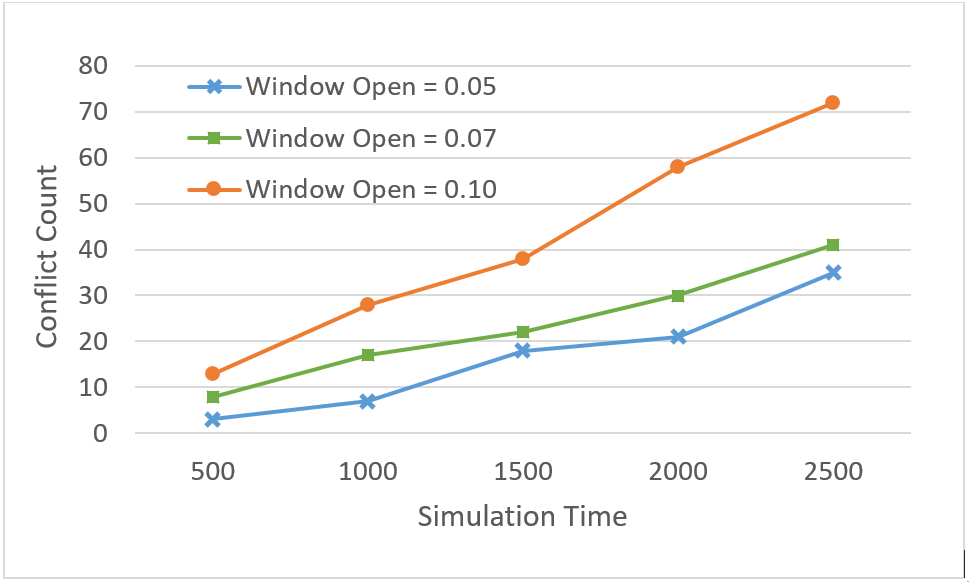} 
        }
    	\subfigure[]{
           \label{A2CountH}
           \includegraphics[scale=0.3,keepaspectratio=true]{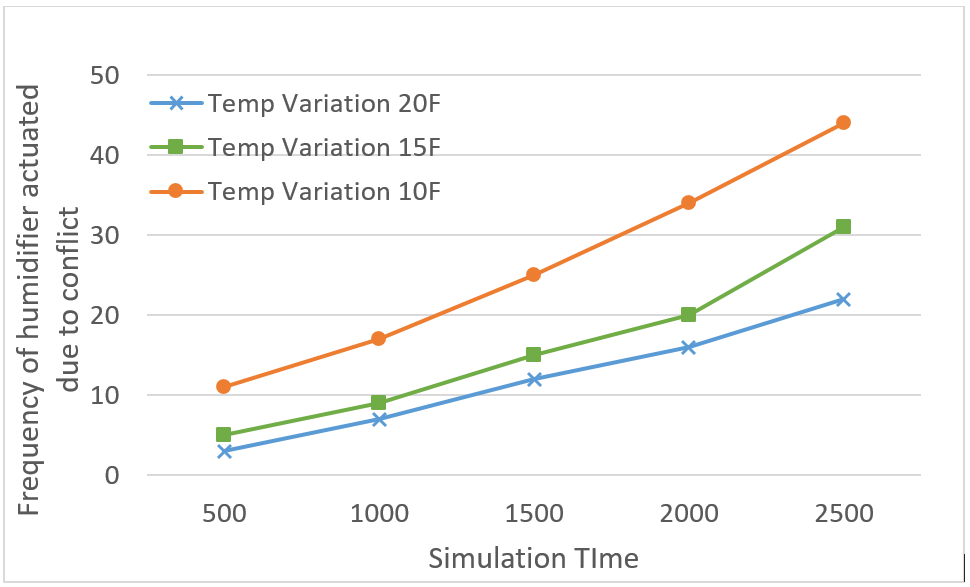} 
        }
        \subfigure[]{
           \label{A1Count}
           \includegraphics[scale=0.3,keepaspectratio=true]{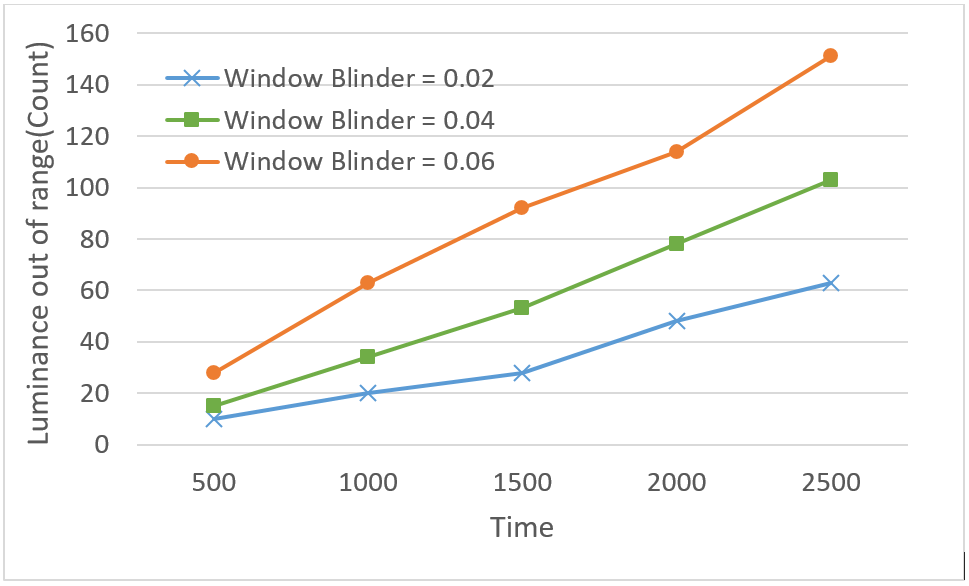} 
        }
    	\subfigure[]{
           \label{A2Count}
           \includegraphics[scale=0.3,keepaspectratio=true]{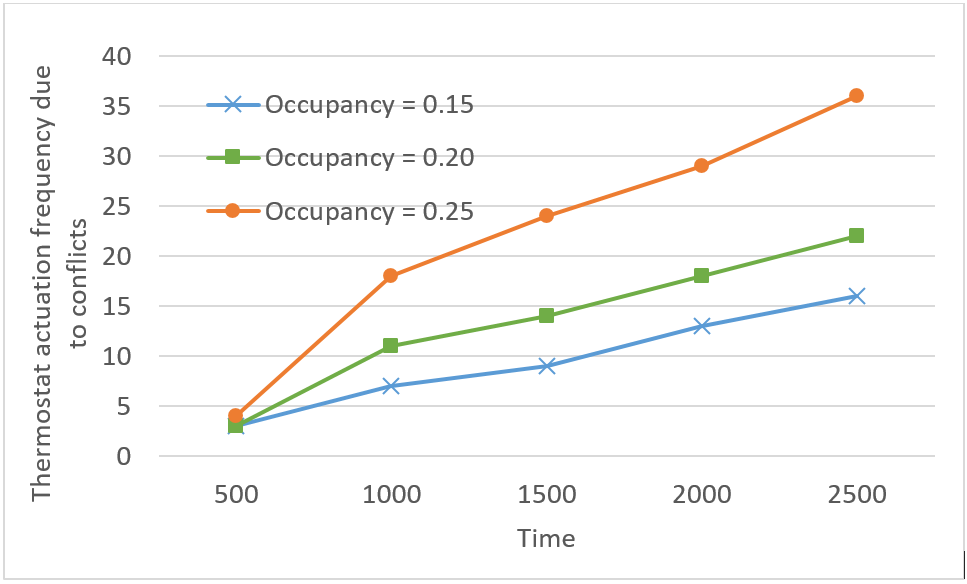} 
        }
        \subfigure[]{
           \label{A4Count}
           \includegraphics[scale=0.3,keepaspectratio=true]{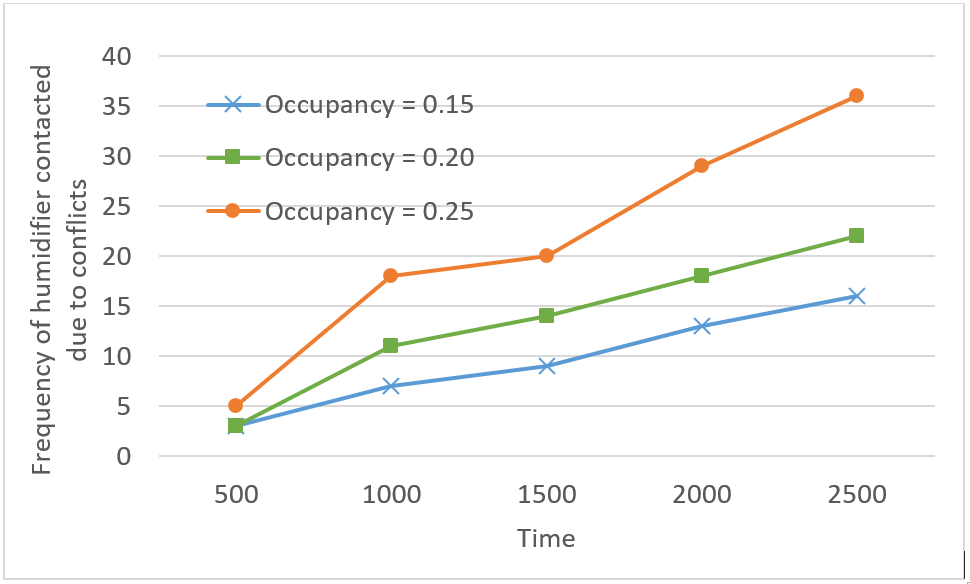} 
        }
	              
       \end{center}
    \vspace{-12pt} 
    \caption{(a) Conflict count when the same alarm is triggered by multiple events, (b) Frequency of thermostat being actuated more than usual due to the window being opened (c) Additional actuation count on the humidifier due to temperature difference in two adjacent rooms (d) Total count of the luminance range exceeding the comfortable range due to conflicts (e) Additional count of the thermostat being actuated due to conflicts between management rules and operational rules, (f) Additional frequency of humidifier being actuated due to conflicts between management rules and random occupancy.}
    \label{ConflictCount}
\vspace{-15pt}    
\end{figure*}

The second experiment measures the change in temperature of a room when a smart window shutter keeps opening at occupant's preference. Like the previous experiment, the window shutter can be opened from the occupant's smart phone or tablet. In addition to that, if carbon monoxide of the house gets increased beyond a certain level, the smart home can be instructed to open the windows immediately for fresh air. This changes the temperature and humidity of the inside of the house. We designed this scenario in our Simulink environment and used the same thermal model of the house~\cite{ThermalSimulink} to get reasonable heat transfer from the outside environment. The results of this experiment are shown in Figure~\ref{C1_Difference}. The blue line indicates the expected temperature reading. However, the red line in this figure indicates how much the temperature reading deviates due to conflicts between the window shutter opening and the thermostat operating rules. This can result in more actuation of the thermostat which will be discussed later in Figure~\ref{C2Count}.

\begin{figure}[htbp]
\centering
\includegraphics[scale=0.34, keepaspectratio=true]{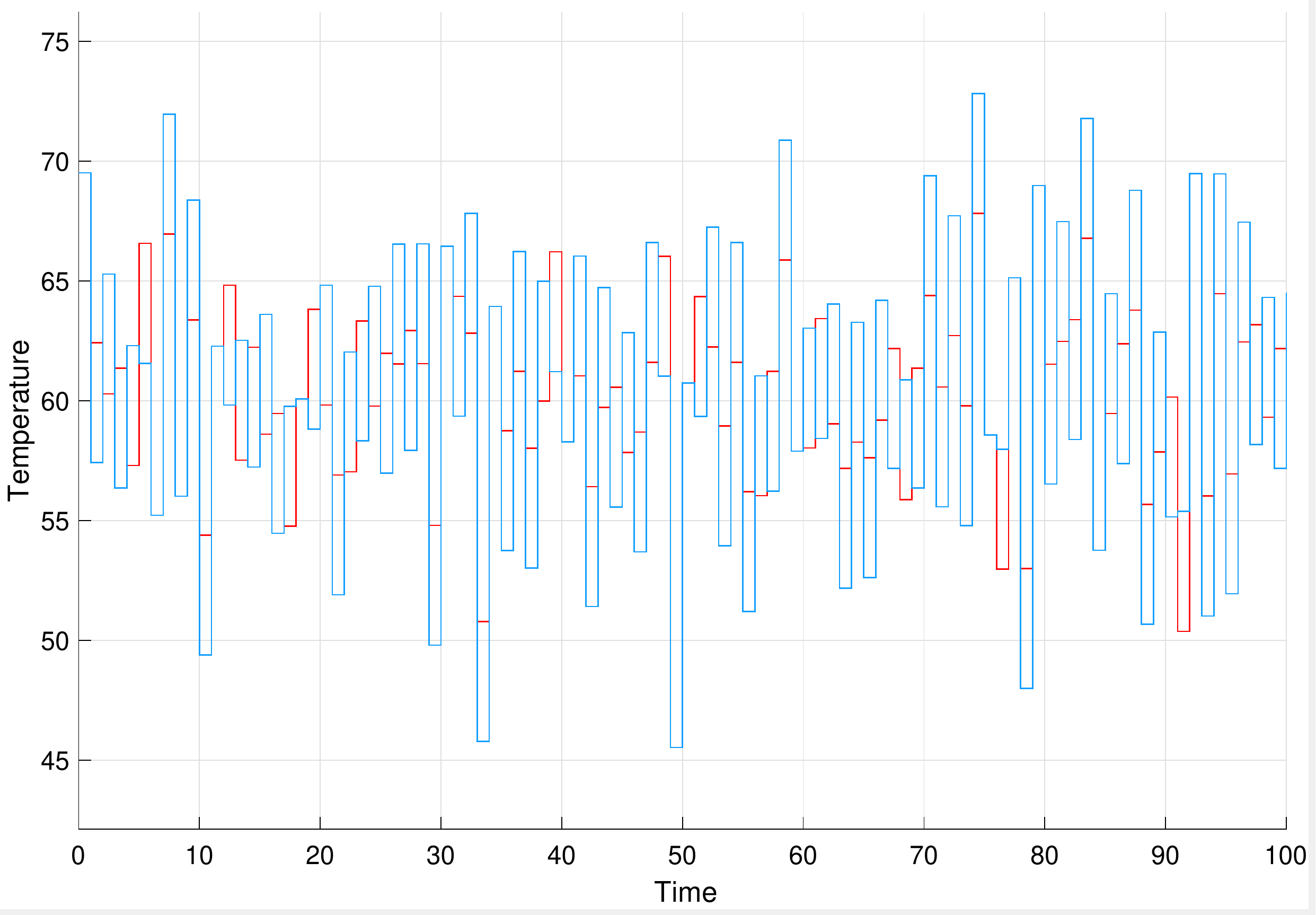}
\vspace{-9pt}
\caption{Effect on temperature of the corridor when it is connected by two rooms of different temperature}
\label{Fig_A1_Difference}
\vspace{-9pt}
\end{figure}

The third experiment captures the impact on temperature of a shared corridor between two rooms. The two rooms are set to have different temperature based on the occupants' preferences. Let us assume the corridor has the thermostat, but no temperature sensor. On the other hand, both rooms have individual sensors. We also assume that room one has the window open which affects the normal temperature of the room by lowering the ambient temperature of room one. On the other hand, room two has no influence from the outside environment, i.e. the window in room two is closed. However, due to two different sensor measurements, the thermostat is instructed to turn on and off more frequently than expected. The temperature, as shown by the red line in Figure~\ref{Fig_A1_Difference}, is the temperature of the corridor, calculated by Simulink. The blue line is the temperature reading during different times from room two. Our intuition says this blue line is supposed to be the temperature reading of the corridor, but it is affected by the temperature from the other room.

\begin{figure}[htbp]
\centering
\includegraphics[scale=0.34, keepaspectratio=true]{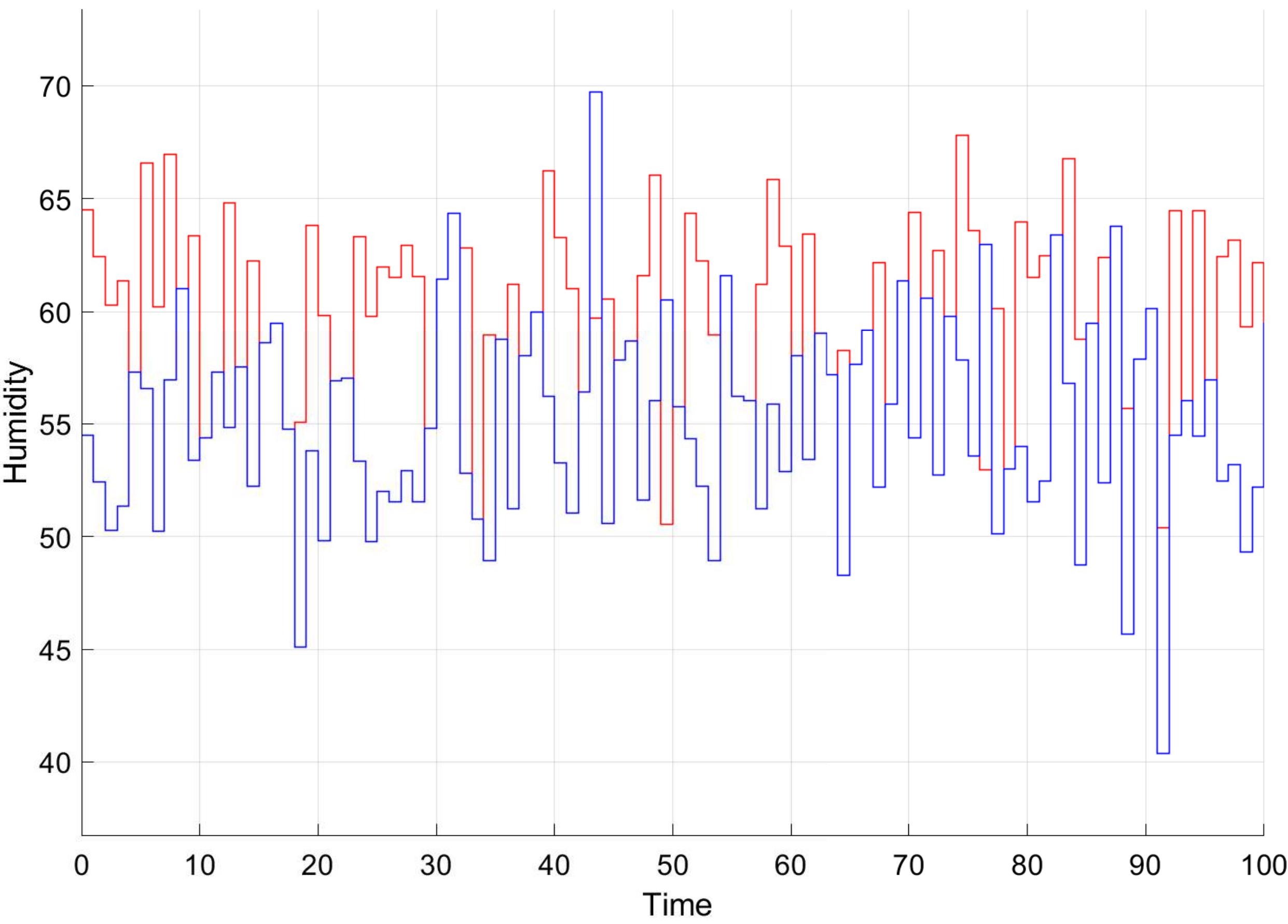}
\vspace{-9pt}
\caption{Effect on Humidity when thermostat and window shutter combinedly changes the temperature and humidity}
\label{A4_Difference}
\vspace{-9pt}
\end{figure}

The fourth experiment characterizes how changes in one environment feature can influence the another environment. Here, we consider the fact that the temperature in a smart home is dependent on thermal radiation and humidity as it is mentioned in~\cite{de2002thermal}. The humidity changes with the temperature of the room. The temperature changes with thermostat and humidifier. Also, the window shutter opening changes the temperature and humidity of the room. The scenario described above is tested and shown in Figure~\ref{A4_Difference}. The expected humidity reading is shown by the blue line, while the real humidity reading is shown by the  red line. Events like the window shutters opening and the humidifier running have caused the temperature of the room to fluctuate. This deviation is overlooked, yet creates discomfort for the occupants and more energy usage due to additional actuation.

To this point, we have conducted our experiments to characterize how different types of conflict can affect the actuators or the environment features. Now, we move our focus on counting the number of conflicts in a given time by varying different parameters. First, we consider the case where the same alarm (actuator) gets triggered when smoke is detected or a water leak is detected. The rules for triggering an alarm in those cases are installed in two different controllers. We consider the probability of smoke detection = 5\% and water leak detection = 7\% in each time unit. The simulation was run for 2000 time units. It is shown in Figure~\ref{C1Count} that the same alarm gets triggered by different events at the same time with the increase of simulation time. Next, we increase both smoke detection and water leak detection to 10\% and find the increment in total conflict count. This type of conflict should be considered as an attacker can compromise only the water system, yet convince the occupants of the smart home that the alarm is due to a fire. That eventually can lead the occupants/target of a smart home to leave the home.

In the next experiment, we count the number of times the window shutters of the smart home are open and the thermostat turns on at the same time. The impact of such conflicts is shown in Figure~\ref{C1_Difference}. As can be in from Figure~\ref{C2Count}, the number of conflicts increases when the window shutter is opened more frequently.

Next, we considered humidity as a dependent feature of temperature. The thermal model of the house is kept the same. The humidity changes based on the temperature which triggers the humidifier. The effect of conflicts in such case are discussed in~\ref{A4_Difference}. First, we run the simulation as if the humidity is not affected by temperature. Next, we re-run the simulation with humidity being dependent on temperature. We count the number of times the humidifier is turned on due to conflicts. As can be seen in Figure~\ref{A2CountH} when there is more temperature variation outside the smart home, the humidifier inside the house gets turned on more often.

We have observed in Figure~\ref{A1_Difference_Luminance} how conflicting actions can affect the same feature. We have seen that the luminance of the room can exceed a comfortable range due to conflicting actions between the smart lights and the smart window blinds. We next count how many times this conflict happens over a specific period of time. We set the probability of the window blinds being opened to 2\%. We then increased this probability while keeping the probability of turning the light on constant (at 10\%) which is reasonably low because if there is an occupant and the window blinds are closed, then the smart lights will turn on immediately. However, the purpose of this experiment is to show that the luminance can get out of range even though there is a low probability of conflicts. Figure~\ref{A1Count} shows that the luminance of the room gets out of range more frequently when there is a greater chance of the window blinds being opened, given that the smart light turned on at the same moment. Similarly, some incidents were observed where a room gets dark when both the window blinds and the smart light turns off.

The next experiment measures the number of conflicts between management rules and operational rules for an IoT system. For a smart building, the management rules may state that the thermostat is not active after 6pm. However, if there are lots of people in a room with their mobile devices or computers on, the temperature of the room will increase. The regular operational rules will then tend to actuate the thermostat to cool the room. Here, we run the simulation for a room in two different ways simultaneously while keeping all parameters same. In the first run, we consider that the occupancy of a room has an effect on temperature and the thermostat is actuated by following basic operational rules. The second run assumes that room occupants and electronic devices does not have a measurable impact on the temperature of the room and that the management rules dictate the operation of the thermostat. Our goal is to observe whether conflicting events cause more actuation of the thermostat. As expected, more actuations of the thermostat are needed when conflicts occur. The building management authorities either overlook this additional actuation, or ignore the comfort of the occupants. The additional number of actuations is shown on Y-axis of Figure~\ref{A2Count}. This number increases when the simulation is executed for a longer time.

The last experiment conducted, is very similar to the previous one. Here, the count of humidifier actuations is measured due to conflicts between operational rules and management rules. We counted the number of additional actuations needed for the humidifier in the smart home. When there are more number of occupants, the air quality degrades and the temperature of the room changes as well. The humidifier turns on to make the environment of the room more comfortable. However, the management rule stipulates not turning on the humidifier after certain time of the day (say 6 pm). There can be conflicts between management rules and regular operational rules based on occupancy of a room. Similar to the previous experiment, we ran simultaneous experiments with different assumptions. During the first run, we assume that room occupancy has negligible impact on the humidity and that the smart building will be operated based on the management rules. For the other run, we make the opposite assumtion, namely that the occupants in the room will have a measurable impact on the humidity of the room. The results are shown in Figure~\ref{A4Count}.

\section{Related Work}

Until today, most of the research efforts have been made for management, efficiency, interoperability, and deployment of IoT systems in the real world. Recently, confidentiality, access control, privacy, and trust issues of IoT technology have been discussed in~\cite{sicari2015security, roman2013features,feng2010study}. The formalization of security properties of an IoT system have rarely been addressed. In IoTSAT~\cite{mohsin2016iotsat}, a formal framework was proposed for security analysis based on device configurations, network topologies, user policies, and IoT-specific attack surface. Recently, the work by~\cite{vannucchi2017symbolic} proposes a verification framework with satisfiable module theory (SMT) for a smart environment with respect to event-condition-action (ECA). However, this research did not propose either safety properties or address conflicts for the smart environment. The work by~\cite{sun2015conflict} made an effort classify conflicts into a set of finite categories. They specified the relation among all building management rules and classified all type of rule conflicts into five categories. Our approach is quite different. Rather than classifying conflicts, we specified safety properties for the components of IoT and the violation of those properties leads to conflicts.

Some preliminary work in the areas of formal modeling and verification for smart home~\cite{corno2014design,coronato2010formal}, intelligent transportation system~\cite{li2013secure}, and health Internet of Things~\cite{kai2013security} has been done. The closest to this work in terms of detecting conflicts are Depsys~\cite{munir2014depsys}, and HomeOS~\cite{dixon2012operating}. Depsys specified and detected conflicts after collecting the functionalities of 35 smart apps used in smart home. It detects the conflicts after they have occurred and in order to solve the situation priorities have been set to the apps so that no two apps can access the same actuator. Our approach, on the other hand, detects conflicts as soon as an event is generated which may immediately cause an action or a set of actions that result in conflicts. We have left the automated resolution of conflicts as future work. 

\section{Conclusion}

The conflicts that are possible in IoT system are often overlooked both in the design phase and during operations. IoT is an automated system and hence the accumulated effects of conflicts on an environment feature or actuator can have more effects initially anticipated. The safety and security of IoT systems is largely dependent of its conflict-less behavior. Hence, the safety properties we formalized in \system consider conflicts as the preeminent threat to the safety and security of IoT system. Furthermore, our model has shown how conflicts can lead to additional actuations which eventually result in more energy consumption. In addition to the contributions mentioned above, we believe our proposed framework will have significant impacts when employed in the \emph{policy monitor} block of \emph{ProvThings}~\cite{wang2018fear}. However, the enforcement of the proposed safety policies for a system is an open challenge. The implementation of an inlined reference monitor (IRM)~\cite{schneider2001language} for enforcing the safety policies of our framework is another interesting research direction. IoT systems are emerging more and more in our daily life, and mechanisms are needed to ensure that these systems are safe, secure, and energy efficient if IoT systems are to be widely deployed and accepted.

\bibliographystyle{IEEEtran}
\bibliography{conflictBib2}

\begin{thebibliography}{10}
\providecommand{\url}[1]{#1}
\csname url@samestyle\endcsname
\providecommand{\newblock}{\relax}
\providecommand{\bibinfo}[2]{#2}
\providecommand{\BIBentrySTDinterwordspacing}{\spaceskip=0pt\relax}
\providecommand{\BIBentryALTinterwordstretchfactor}{4}
\providecommand{\BIBentryALTinterwordspacing}{\spaceskip=\fontdimen2\font plus
\BIBentryALTinterwordstretchfactor\fontdimen3\font minus
  \fontdimen4\font\relax}
\providecommand{\BIBforeignlanguage}[2]{{%
\expandafter\ifx\csname l@#1\endcsname\relax
\typeout{** WARNING: IEEEtran.bst: No hyphenation pattern has been}%
\typeout{** loaded for the language `#1'. Using the pattern for}%
\typeout{** the default language instead.}%
\else
\language=\csname l@#1\endcsname
\fi
#2}}
\providecommand{\BIBdecl}{\relax}
\BIBdecl

\bibitem{dave2011next}
E.~Dave \emph{et~al.}, ``How the next evolution of the internet is changing
  everything,'' \emph{The Internet of Things}, 2011.

\bibitem{IoTDDOS}
\BIBentryALTinterwordspacing
S.~COBB, ``10 things to know about the october 21 iot ddos attacks,'' 2016.
  [Online]. Available:
  \url{http://www.welivesecurity.com/2016/10/24/10-things-know-october-21-iot-ddos-attacks/}
\BIBentrySTDinterwordspacing

\bibitem{de2002thermal}
R.~J. De~Dear and G.~S. Brager, ``Thermal comfort in naturally ventilated
  buildings: revisions to ashrae standard 55,'' \emph{Energy and buildings},
  vol.~34, no.~6, pp. 549--561, 2002.

\bibitem{TriskaProlog}
M.~Triska, ``Theorem proving with prolog,''
  \url{https://www.metalevel.at/prolog/theoremproving}, accessed: 2018-07-27.

\bibitem{wielemaker2012swi}
J.~Wielemaker, T.~Schrijvers, M.~Triska, and T.~Lager, ``Swi-prolog,''
  \emph{Theory and Practice of Logic Programming}, vol.~12, no. 1-2, pp.
  67--96, 2012.

\bibitem{ThermalSimulink}
M.~Simulink, ``Thermal model of a house,''
  \url{https://www.mathworks.com/help/simulink/examples/thermal-model-of-a-house.html},
  accessed: 2018-07-27.

\bibitem{sicari2015security}
S.~Sicari, A.~Rizzardi, L.~A. Grieco, and A.~Coen-Porisini, ``Security, privacy
  and trust in internet of things: The road ahead,'' \emph{Computer Networks},
  vol.~76, pp. 146--164, 2015.

\bibitem{roman2013features}
R.~Roman, J.~Zhou, and J.~Lopez, ``On the features and challenges of security
  and privacy in distributed internet of things,'' \emph{Computer Networks},
  vol.~57, no.~10, pp. 2266--2279, 2013.

\bibitem{feng2010study}
H.~Feng and W.~Fu, ``Study of recent development about privacy and security of
  the internet of things,'' in \emph{Web Information Systems and Mining (WISM),
  2010 International Conference on}, vol.~2.\hskip 1em plus 0.5em minus
  0.4em\relax IEEE, 2010, pp. 91--95.

\bibitem{mohsin2016iotsat}
M.~Mohsin, Z.~Anwar, G.~Husari, E.~Al-Shaer, and M.~A. Rahman, ``Iotsat: A
  formal framework for security analysis of the internet of things (iot),'' in
  \emph{Communications and Network Security (CNS), 2016 IEEE Conference
  on}.\hskip 1em plus 0.5em minus 0.4em\relax IEEE, 2016, pp. 180--188.

\bibitem{vannucchi2017symbolic}
C.~Vannucchi, M.~Diamanti, G.~Mazzante, D.~Cacciagrano, R.~Culmone,
  N.~Gorogiannis, L.~Mostarda, and F.~Raimondi, ``Symbolic verification of
  event--condition--action rules in intelligent environments,'' \emph{Journal
  of Reliable Intelligent Environments}, vol.~3, no.~2, pp. 117--130, 2017.

\bibitem{sun2015conflict}
Y.~Sun, X.~Wang, H.~Luo, and X.~Li, ``Conflict detection scheme based on formal
  rule model for smart building systems,'' \emph{Human-Machine Systems, IEEE
  Transactions on}, vol.~45, no.~2, pp. 215--227, 2015.

\bibitem{corno2014design}
F.~Corno and M.~Sanaullah, ``Design-time formal verification for smart
  environments: an exploratory perspective,'' \emph{Journal of Ambient
  Intelligence and Humanized Computing}, vol.~5, no.~4, pp. 581--599, 2014.

\bibitem{coronato2010formal}
A.~Coronato and G.~De~Pietro, ``Formal design of ambient intelligence
  applications,'' \emph{Computer}, vol.~43, no.~12, pp. 60--68, 2010.

\bibitem{li2013secure}
S.~Li, P.~Gong, Q.~Yang, M.~Li, J.~Kong, and P.~Li, ``A secure handshake scheme
  for mobile-hierarchy city intelligent transportation system,'' in
  \emph{Ubiquitous and Future Networks (ICUFN), 2013 Fifth International
  Conference on}.\hskip 1em plus 0.5em minus 0.4em\relax IEEE, 2013, pp.
  190--191.

\bibitem{kai2013security}
K.~Kai, Z.-b. PANG, and W.~Cong, ``Security and privacy mechanism for health
  internet of things,'' \emph{The Journal of China Universities of Posts and
  Telecommunications}, vol.~20, pp. 64--68, 2013.

\bibitem{munir2014depsys}
S.~Munir and J.~A. Stankovic, ``Depsys: Dependency aware integration of
  cyber-physical systems for smart homes,'' in \emph{Cyber-Physical Systems
  (ICCPS), 2014 ACM/IEEE International Conference on}.\hskip 1em plus 0.5em
  minus 0.4em\relax IEEE, 2014, pp. 127--138.

\bibitem{dixon2012operating}
C.~Dixon, R.~Mahajan, S.~Agarwal, A.~Brush, B.~Lee, S.~Saroiu, and P.~Bahl,
  ``An operating system for the home,'' in \emph{Proceedings of the 9th USENIX
  conference on Networked Systems Design and Implementation}.\hskip 1em plus
  0.5em minus 0.4em\relax USENIX Association, 2012, pp. 25--25.

\bibitem{wang2018fear}
Q.~Wang, W.~U. Hassan, A.~Bates, and C.~Gunter, ``Fear and logging in the
  internet of things,'' in \emph{ISOC NDSS}, 2018.

\bibitem{schneider2001language}
F.~B. Schneider, G.~Morrisett, and R.~Harper, ``A language-based approach to
  security,'' in \emph{Informatics}.\hskip 1em plus 0.5em minus 0.4em\relax
  Springer, 2001, pp. 86--101.

\end{thebibliography}

\end{document}